\documentclass[manuscript]{aastex} %single column double-spaced document
\shorttitle{FLARE INDUCED SEISMICITY IN ACTIVE REGION AND RELATED GLOBAL WAVES IN THE SUN}
\shortauthors{Brajesh Kumar et al.}

\begin{document}

\title{ON THE FLARE INDUCED SEISMICITY IN THE ACTIVE REGION NOAA 10930 AND RELATED ENHANCEMENT OF GLOBAL WAVES IN THE SUN}

\author{BRAJESH KUMAR}
\affil{Udaipur Solar Observatory, Physical Research Laboratory,
 Dewali, Badi Road, Udaipur 313 004, India}
\email{brajesh@prl.res.in}
\and
\author{P. VENKATAKRISHNAN}
\affil{Udaipur Solar Observatory, Physical Research Laboratory, Dewali, Badi Road,
Udaipur 313 004, India}
\email{pvk@prl.res.in}
\and
\author{SAVITA MATHUR}
\affil{High Altitude Observatory, 3080 Center Green Drive, Boulder, CO 80302, USA}
\email{savita@ucar.edu}
\and
\author{SANJIV KUMAR TIWARI}
\affil{Max-Planck-Institut f\"{u}r Sonnensystemforschung, Max-Planck-Str. 2,
37191 Katlenburg-Lindau, Germany}
\email{tiwari@mps.mpg.de}
\and
\author{R. A. GARC\'IA}
\affil{Laboratoire AIM, CEA/DSM-CNRS, Universit\'e Paris 7 Diderot, IRFU/SAp,
Centre de Saclay, 91191, Gif-sur-Yvette, France}
\email{rafael.garcia@cea.fr}

%\slugcomment{Not to appear in Nonlearned J.}
\begin{abstract}
A major flare (of class X3.4) occurred on 13 December 2006 in the active region NOAA 10930. This flare
event has remained interesting for solar researchers for studies related to particle acceleration
during the flare process and the reconfiguration of magnetic fields as well as fine scale features
in the active region. The energy released during flares is also known to induce acoustic
oscillations in the Sun. Here, we analyze the line-of-sight velocity patterns in this active region
during the X3.4 flare using the Dopplergrams obtained by Global Oscillation Network Group (GONG) instrument.
We have also analyzed the disk-integrated velocity observations of the Sun obtained by Global
Oscillation at Low Frequency (GOLF) instrument onboard {{\em SOHO (Solar and Heliospheric
Observatory)}} spacecraft as well as full-disk collapsed velocity signals from GONG observations
during this flare to study any possible connection between the flare
related changes seen in the local and global velocity oscillations in the Sun. We apply wavelet
transform to the time series of the localized velocity oscillations as well as the global velocity
oscillations in the Sun spanning the flare event. The line-of-sight velocity shows significant
enhancement in some localized regions of the penumbra of this active region during the flare. The
affected region is seen to be away from the locations of the flare ribbons and the hard X-ray
footpoints. The sudden enhancement in this velocity seems to be caused by the Lorentz force driven
by the ``magnetic jerk'' in the localized penumbral region. Application of wavelet analysis to these
flare induced localized seismic signals show significant enhancement in the high-frequency domain
(5$<$$\nu$$<$8~mHz) and a feeble enhancement in the $p$-mode oscillations (2$<$$\nu$$<$5~mHz) during
the flare. On the other hand, the wavelet analysis of GOLF velocity data and the full-disk
collapsed GONG velocity data spanning the flare event
indicate significant post-flare enhancements in the high-frequency global velocity oscillations in
the Sun as evident from the Wavelet Power Spectrum and the corresponding scale-average variance.
The present observations of the flare induced seismic signals in the active region in context to
the driving force is different as compared to the previous reports on such cases. We also find indications
of a connection between flare induced localized seismic signals and the excitation of global high-frequency
oscillations in the Sun.
\end{abstract}


\section{INTRODUCTION}

Solar flares release a large amount of energy in the solar environment and produce energetic particles
moving with relativistic speeds. The bremsstrahlung radiation generated by deceleration of the particles
while striking the target (chromospheric material) produces hard X-ray emission \citep{brow71}, while
the gyrosynchrotron emission due to motion of charged particles along magnetic fields produces
microwave radiation \citep{kaki62, huds72} and $\gamma$-rays. Apart from these effects, the mechanical
response of the solar atmosphere to flares has also been reported \citep[and references therein]{kumar10}.
In particular, the observations of
\cite{venkat08} show co-spatial evolution of the photospheric Doppler enhancements and the chromospheric
H$\alpha$ flare ribbons in the active region NOAA 10486 during the 4B/X17.2 class solar flare of 28
October 2003. In addition, a feeble enhancement in global high-frequency velocity oscillations was seen
to be induced by this flare and two other flares \citep{kumar10}. A further opportunity to examine this effect
was seen in active region NOAA 10930 which appeared on the solar disk during December 2006 and produced a
lot of space weather related activity \citep{li09,ning08}. The most intense flare (of class X3.4) in
this active region was reported on 13 December 2006. The X-ray Telescope (XRT; \cite{golu07}) on board
{\em Hinode} spacecraft \citep{kosu07} observed the evolution of sheared coronal magnetic fields
\citep{su07} during this flare. The EUV Imaging Spectrometer (EIS; \cite{culhane07}) measured plasma
flows \citep{imad07,asai08} and turbulent motions \citep{imad08} during the flare. In addition, some
interesting photospheric and chromospheric disturbances were also reported in this active region during
the flare \citep{mino09}. The Solar Optical Telescope (SOT; \cite{tsun08}) on board {\em Hinode}
\citep{kosu07} spacecraft has obtained high-resolution photospheric and chromospheric observations of
this flare event in G-band (4305~\AA) and Ca~{\sc ii}~H (3968~\AA), respectively. These observations
have shown elongated flare ribbons which moved apart as the flare progressed with time
\citep{isob07, jing08}.
The hard X-ray (HXR) images of this flare event captured by {\em RHESSI} spacecraft show double-footpoint
HXR sources located on the flare ribbons \citep{su07,mino09}. The {\em Hinode} observations of this
active region have also shown lateral motion of penumbral filaments in the active region during the
flare \citep{gosain09} as well as interesting changes in the magnetic field inclinations \citep{gosain10}
and the net current \citep{ravindra11}.
\cite{koso07} have revealed high-frequency oscillations excited by this flare in the sunspot chromosphere
using the Ca~{\sc ii}~H intensity images obtained by SOT on board {\em Hinode} spacecraft.

In this paper, we present a detailed study of the flare induced seismic signals in the active region
NOAA 10930 during the flare using the full-disk Dopplergrams obtained by the Global Oscillation Network Group
instrument (GONG; \cite{harv95}). A preliminary report on the flare related localized enhancements
seen in this active region during this flare event has been communicated by \cite{kumar11}. We have
used wavelet techniques
\citep{torr98,liu07} to analyze the velocity time series. This gives us an opportunity to examine whether
there are any short-lived pulsations induced by the flare. An important feature of
the wavelet analysis is that we can see the evolution of any physical parameter in the frequency
domain as a function of time. Recently, it has been shown
that flares can also influence high-frequency (5--8~mHz) global acoustic oscillations
in the Sun \citep{karoff08, kumar10}. Motivated by these findings, we have also looked for enhancements of
the global velocity oscillations in the Sun for this X3.4 class flare of 13 December 2006 using the
disk-integrated velocity observations of the Sun obtained from Global Oscillation at Low Frequency
(GOLF; \cite{gabr95}) on board {\em SOHO (Solar and Heliospheric Observatory)} spacecraft \citep{domi95}
as well as full-disk collapsed velocity signals from GONG observations.
We have applied wavelet analysis to these GOLF velocity data and the full-disk collapsed velocity
data from GONG spanning the flare event. The idea behind
this is to study any possible connection between the flare-induced enhancements in the local velocity
oscillations and the global velocity oscillations in the Sun.

\section{THE OBSERVATIONAL DATA}

The active region NOAA 10930 was one of the largest group of sunspots observed during the minimum phase
of the solar activity cycle 23. It appeared on the east limb of the Sun on 7 December 2006 in the
southern hemisphere. This complex active region belonged to $\beta\gamma\delta$ class and produced
several X-, M- and C-class flares during its passage on the solar disk. On 13 December 2006, a major flare
(of class X3.4) occurred in this active region when it was located at S06W27. This was one of the
largest flares that occurred during the solar minimum of the activity cycle 23 and produced high-speed
halo Coronal Mass Ejection (CME) and extremely energetic emissions \citep{wang08b,li09}.  This flare was
well observed by Nobeyama Radio Polarimeters (NoRP), {\em RHESSI}, {\em GOES}, {\em Hinode}, GONG and
other ground- or space-based instruments. Although {\em RHESSI} missed the first impulsive phase of
the flare in hard X-ray emissions, the full event was covered in microwave emissions by NoRP and
in soft X-ray emission by {\em GOES}. We have used data from GONG and GOLF for the analysis of
velocity oscillations during the flare. The details of the data are given as follows.

\subsection{GONG DATA}

The GONG observations comprise of solar full-disk photospheric Dopplergrams taken in
Ni~{\sc i}~6768~\AA~line.
These full-disk Dopplergrams are obtained once a minute with a spatial sampling rate of $\sim$2.5~arcsec
per pixel. In this study, we have used the calibrated and solar-rotation signal removed GONG Dopplergrams
between 01:00~UT and 03:02~UT taken on 13 December 2006 spanning the flare event. The active region
NOAA 10930 was tracked using heliographic coordinates in the full-disk Dopplergrams,
which were interpolated to a 1~arcsec~pixel size. We selected a grid of 60$\times$60 pixels
($\sim$150$\times$150~arcsec) covering this active region for preparing the tracked data. The tracked
velocity images were further co-aligned with a sub-pixel accuracy. We are interested in examining
the effect of the flare on the line-of-sight velocity signals much beyond the impulsive phase
of the flare ($\sim$02:25~UT) as observed in microwave emissions by NoRP and the maximum of
soft X-ray flux at $\sim$02:35~UT as reported by {\em GOES}. However, there
was a disruption in the GONG observations after 03:02~UT and hence we have analyzed the GONG data
only till 03:02~UT time. In the Figure~1(a), we show the continuum intensity map of the active
region NOAA 10930 obtained by the GONG instrument on 13 December 2006 at 02:26~UT. The field-of-view
of this intensity map is the same as that for the aforementioned rasters of co-aligned velocity images.

\subsection{GOLF DATA}

The GOLF observations comprise of disk-integrated photospheric line-of-sight velocity of the Sun
using Na~D~lines. These observations are obtained at a cadence of 10~s. However, we have used rebinned
velocity data from GOLF for every 60~s to match with the GONG observations. Before the beginning of
`{\em SOHO} vacation' in June 1998 and since December 2002 the GOLF experiment has been probing the
blue wing of the Na~D~lines
at two points separated by $\sim$0.0003~nm and displaced by $\sim$0.0108~nm from the centers of the
lines \citep{garc05}. Hence, the GOLF observations are similar in nature to other known sun-as-a-star
velocity measurements \citep{palle99}, such as, IRIS and BiSON, and these observations are mostly
confined to the solar photosphere but with a small contribution of the chromosphere which depends
on the orbital velocity \citep{jim07, jim05}. Therefore, the GOLF observations are also
comparable with GONG observations with respect to the observing heights in the solar photosphere
\citep{jim07}. In this study, we have used the GOLF observations for the period from 01:00~UT to 04:15~UT
on 13 December 2006, which covers the GONG observations used in our analysis. In fact, there was a
disruption in GOLF
observations after 04:15~UT and hence we could not use these velocity observations beyond this time. We
have worked with the standard velocity time series \citep{garc05,ulri00a} from GOLF. The GOLF housekeeping
data have sometimes shown anomalous behaviour during major flares. We have checked and confirmed
that the housekeeping data corresponding to our velocity time series are free from such effects. A
two-point backward difference filter \citep{garc08} is applied to the velocity series to remove the
effect of the orbital motion of the satellite with respect to the Sun.

\section{ANALYSIS AND RESULTS}

The sequence of the co-aligned grids ($\sim$150$\times$150~arcsec) from the GONG Dopplergrams covering
the active region NOAA 10930 are analyzed to see the effect of this flare event on the local
line-of-sight velocity signals in different locations of the active region. Simultaneously,
we analyze the velocity time
series from GOLF to examine the effect of this flare on global velocity oscillations in the Sun.
As mentioned earlier, we apply wavelet transform to these data to investigate the effect of the
flare on the solar velocity oscillations. Wavelet techniques have the capability to detect episodic
and weak pulsations in a given time series \citep{torr98}. In the application of wavelet techniques,
here we have limited our analysis to the periods of less than 25\% of the time series length. This
ensures a better reliability of the periods and technically this is known as ``cone of influence''.
Here, the cone of influence takes into account the fact that we should observe at least four
periodicities to be confident that it is a real signal. This gives the limit in frequency above
which any periodicity is reliable. For example, given a time series of two hours, we should be able to
detect periods $<$ 30~min. However, the frequency resolution of the wavelet transform is still
decided by the temporal resolution in our velocity time series, which is 60~s. In our wavelet
analysis, we have estimated two confidence levels of detection corresponding to the probability
of 90\% and 50\% . This helps in deciding the episodes when wavelet power is above noise. In the
Wavelet Power Spectrum, the confidence levels corresponding to 90\% and 50\% are outlined using
contour maps.

\subsection{ANALYSIS OF VELOCITY SIGNALS IN THE ACTIVE REGION USING GONG DOPPLERGRAMS}

In order to understand the evolution of flare induced seismic waves in the Sun, we also require
the information related to time evolution of high energy radiations from the Sun during the flare.
As mentioned earlier, {\em RHESSI} missed the first impulsive phase of the flare in hard X-ray
emissions. On the other hand, NoRP fully covered this flare in microwave emissions and simultaneously
the {\em GOES} spacecraft has observed this event in soft X-ray emissions from the Sun before, during
and after the flare. For this flare event, the flare intensity was highest at 02:25~UT as seen in
microwave emission observations from NoRP \citep{ning08,mino09}. However, the soft X-ray observations
from {\em GOES} show peak intensity around 02:35~UT. The difference in the peak time of the flare
as observed by NoRP and {\em GOES} could be explained as follows. The microwave flare observations
represent the energy released as a proxy for the hard X-ray observations while the soft X-ray
observations track the thermal emission from the hot plasma that is evaporated into the coronal
loops as a result of the energy release. Since the microwave emission profile during the flare
has been already shown
by \cite{ning08} and \cite{mino09}, here we plot the time evolution of {\em GOES} soft X-ray
emissions from the Sun during the flare in the top panel of Figure~2. Getting back to the analysis
of velocity signals in the active region during the flare, we have computed a
{\em root mean square (rms)} velocity map from the tracked GONG Dopplergrams for the period between
02:00~UT and 03:02~UT and for the same field-of-view as shown in the Figures~1(a). The corresponding
surface plot of this {\em rms} map is shown in the Figure~1(c). The Figure~1(b) clearly shows a
relatively big bright patch (indicated by `K' in this image) in the
localized penumbral region of the active region, which is the site of flare induced
large velocity flows. We have also marked another bright region `P' in the penumbra
(diagonally opposite to `K'). `K' and `P' are locations of peaks which are higher than 10$\sigma$,
where $\sigma$ is the {\em rms} of the spatial distribution of the values displayed in the
Figures~1(b)~and~1(c). We have also chosen a region `Q' in the quiet region (far away from the
active region) for comparison. We have analyzed the mean velocity
flows in grids of 3$\times$3 pixels surrounding the centroids of the regions `K', `P', and `Q'.
In the Figure~2, we show the temporal evolution of line-of-sight velocity signals obtained for the
regions `K', `P' and `Q', respectively with respect to the evolution of {\em GOES} soft X-ray flux
during the flare. We note that the line-of-sight velocity of the region `K' evolves from
a downflow of $\sim$500~m/s at $\sim$~01:00~UT and changes into an upflow at $\sim$02:00~UT. It
shows transient enhancement at $\sim$02:26~UT, during the main impulsive phase of the flare,
and thereafter decreases to $\sim$100~m/s around $\sim$02:45~UT. This evolution is modulated by
an oscillatory behavior. The region `P' shows a more gradual evolution from an upflow of $\sim$200~m/s
at $\sim$01:00~UT to a downflow of $\sim$200 m/s at $\sim$02:00~UT and finally reverts back to the
original upflow after the flare. The oscillatory modulation is seen for this case also. The region `Q'
shows the normal solar oscillations about a zero mean. The Wavelet Power Spectrum (WPS)
and Global-wavelet Power Spectrum (GWPS) obtained for the regions `K', `P'
and `Q', respectively, are shown in the Figures~3(a), 4(a), and 5(a). The GWPS is a collapsogram
of the WPS along time. The Figures~3(b), 4(b) and 5(b) show the scale-average
variance obtained from the corresponding WPS in the frequency regimes: (2--5~mHz), and
(5--8~mHz), respectively in the top and the bottom panels. Basically, these are collapsograms of the WPS along
the frequency of the wavelet in the chosen range. For this quantity, we have calculated the significance
levels for 50\% and 90\% probability. These results indicates major enhancement in high-frequency
(5--8~mHz) velocity oscillations in the region `K' at about 02:30~UT which is around the impulsive
phase of the flare. In the frequency regime (2--5~mHz) where the $p$ modes are expected, the
power is suppressed in the penumbral regions `K' and `P' (top panels of Figure~3(b) and 4(b)), which
is attributed to the absorption of $p$ modes by magnetic field concentrations in these penumbral
regions. Whereas, in a quiet region `Q' of the Sun, the power of the $p$ modes
is above 50\% and 90\% significance levels (top panel of Figure~5(b)). Finally, at high frequency
(5--8~mHz), significant increase of power after
the flare is observed in the average variance (bottom panel of Figure~3(b)) for the region `K'. We do
observe a feeble increase (less than 50\% significance level) in power in the frequency regime (2--5~mHz)
during the flare in the region `K' while no flare related enhancements are observed for the regions `P'
and `Q'.

\subsection{ANALYSIS OF GLOBAL VELOCITY OSCILLATIONS USING GOLF VELOCITY DATA}

The temporal evolution of the filtered velocity signals from GOLF instrument between 01:00~UT and
04:15~UT spanning the flare event of 13 December 2006 is shown in the Figure~2 (second panel from the top).
These velocity oscillations show a relative increase in the amplitude of velocity oscillations after
the flare. Wavelet transform is applied to the time series of these velocity signals. The panels in Figure~6(a)
show the WPS and GWPS obtained from the aforementioned velocity time series from GOLF. Here, we
observe enhanced high-frequency power in the global-velocity oscillations in the Sun after the flare.
In the Figure~6(b), we show the scale-average variance in the frequency regimes: (2--5~mHz)
and (5--8~mHz), respectively in the top and the bottom panels, obtained from the WPS for GOLF data as
shown in the Figure~2. A close inspection of the scale-average variance in the high-frequency regime
reveals that there is a small enhancement in high-frequency pulses during the impulsive phase of
the flare ($\sim$02:30~UT), then it drops down and again depicts major pulses with confidence
level more than 90\% beginning 03:30~UT
(an hour later to the flare maximum). We conjecture that the enhanced velocity pulses seen during the
flare is the contribution from the chromospheric oscillations, whereas the major enhancement in the
velocity pulses seen several minutes later to the flare is contribution from photospheric oscillations.
This could be understood as follows. According to \cite{cessa10}, the GOLF observations will be limited
to photospheric heights in the quiet Sun, however these observations will become sensitive to
chromospheric heights during the flare. Hence, just after the flare, the pulses seen in the GOLF data
could be a contribution from the chromospheric oscillations. These oscillations could be
interpreted as a narrowband oscillatory response
of the chromospheric acoustic resonator to a broadband impulsive excitation as suggested by
\cite{botha11}. However, the strong pulses observed by the
GOLF much later to the flare should be from the photospheric oscillations as the GOLF observations
would return to normal mode. Here, the presence of the pulses corresponding to the enhanced chromospheric
oscillations is also consistent with the chromospheric umbral oscillations seen by \cite{koso07} in
the SOT/{\em Hinode} Ca~{\sc ii}~H data during this flare event.

\subsection{ANALYSIS OF GLOBAL VELOCITY OSCILLATIONS USING GONG VELOCITY DATA}

The sequence of GONG full-disk Dopplergrams obtained during 01:00-03:02~UT spanning the flare event
is subjected to the two-point backward difference filter to enhance the velocity signals from
$p$ modes and high-frequency waves above the solar background. These filtered Dopplergrams are
then collapsed (excluding the limb pixels to avoid noise) to a single velocity value, which
should represent the global modes of oscillations in the Sun \citep{kumar10}. The analysis of
these velocity signals would give us the opportunity to compare the behaviour of flare induced
global waves in the Sun at different heights in the solar atmosphere. The temporal
evolution of these filtered full-disk collapsed velocity signals is shown in the Figure~2
(third panel from the top). Here, we observe that these velocity signals are dominated by
normal $p$-mode oscillations similar to those seen in the time series for the Q-region as
shown in the Figure~2 (sixth panel from the top). In the Figure~7(a), we show the WPS and
GWPS obtained from the
aforementioned velocity time series from GONG. Here, we do observe enhanced high-frequency
power in the global-velocity oscillations in the Sun after the flare. In the Figure~7(b),
we show the scale-average variance in the frequency regimes: (2--5~mHz)
and (5--8~mHz), respectively in the top and the bottom panels, obtained from the WPS as shown
in the Figure~7(a). The scale-average variance in the high-frequency regime indicates
enhancement in high-frequency pulse after the impulsive phase of the flare. However, the
epoch of this enhancement is different from that seen in the GOLF data. This could be attributed
to the different observing heights of these two instruments and their behaviour during the
flare. As mentioned in the above Section, GOLF observations would get affected during the
flare, however the GONG observations are expected to be maintained in the same height in
the solar atmosphere during the flare.

\section{DISCUSSION AND CONCLUSIONS}

The flare event of 13 December 2006 that occurred in the active region NOAA 10930 has remained very
interesting for the solar researchers as it has provided an opportunity to understand various
inter-linked physical processes taking place in the different layers of the Sun, from photosphere to
the corona. Also, there is an indication of a connection between the flare induced local and
global oscillations in the Sun. The impact of this flare (of class X3.4) was very much pronounced
on the solar photosphere and chromosphere. The G-band (4305~\AA) photospheric images obtained by
SOT on board {\em Hinode} spacecraft have shown flare ribbons and lateral motion of penumbral
filaments during the flare. The chromospheric observations taken by SOT in Ca~{\sc ii}~H (3968~\AA)
show sustained high-frequency umbral oscillations induced by this flare \citep{koso07}. The
photospheric velocity
observations from GONG also show enhanced high-frequency seismic signals in the localized
regions of the active
region following the impulsive phase of the flare. \cite{sych09} have reported bursts of microwave
emission with 3 minute intervals during several flares which they relate to the photospheric
3 minute oscillations in the active regions. It is quite possible that the sustained chromospheric
high-frequency umbral oscillations as seen by \citep{koso07} are also related to the enhanced
high-frequency oscillations in the localized regions of the active region as seen in the GONG data.
The location of the velocity change is stationary and away from the G-band flare ribbons
(c.f. \cite{mino09}). This is in contrast to the sites of
velocity changes reported for the 28 October 2003 flare by \cite{venkat08}, which were located in
the photospheric Doppler ribbons
and were seen to move co-spatially with the H$\alpha$ flare ribbons. The earlier studies have reported
seismic emissions either at the location of moving hard X-ray footpoints or in the H$\alpha$ flare
kernels \citep[and references therein]{venkat08}. Therefore, their observations could be explained
by the possible physical processes like: (i) chromospheric shocks propagating through the photosphere
and into the solar interior \citep{fisher85}, or (ii) high-energy particle beam impinging on the solar
photosphere \citep{zhar07}. However, the present observation is a new kind of photospheric response to a
solar flare. Therefore, it requires a different mechanism to produce the seismic response. The
Lorentz force generated by the magnetic field changes has been shown to be large enough to power
seismic waves \citep{huds08,pet10}. \cite{huds08} have estimated that the Lorentz forces of the size
$\sim$$10^{22}$~dynes could be responsible for generating seismic waves. It is worth noting that we do
see sudden line-of-sight magnetic field changes ($\sim$85~G) as obtained from GONG magnetograms in the
affected region `K' (mean magnetic field $\sim$1200~G) that could result in Lorentz force
($B_z \cdot\delta B_z/4\pi$) of the required order ($\sim$$10^{22}$~dynes) over the area of the region
`K' ($\sim$$10^{18}~cm^{2}$). We have also examined whether the magnetic fluctuations in the
penumbral region `K' are the result of any serious cross-talk from the velocity fluctuations in
this region and found no evidence for this. However, our estimate suffers from the lack of
vector magnetic field data spanning the flare event. In this context, it is important to mention
the findings of \cite{brown92}. Their observations
indicate that there are localized patches of magnetic flux which act as sources of high-frequency
oscillations. This conclusion was based on spatial maps of acoustic power distribution and
magnetograms. However, they had not looked at the transient effects. On the other hand, we report
on localized enhancement in line-of-sight velocities in the active region related to a flare and
these flare induced seismic signals have significant high-frequency oscillations. But in both
cases, the magnetic field seems to be the driving force behind these high-frequency
oscillations.

The results of wavelet analysis of GOLF velocity data spanning the flare event are shown in the
Figures~6(a) and 6(b). As mentioned earlier, we could use the GOLF data for the period from 01:00~UT
to 04:15~UT on 13 December 2006, which covers the GONG observations of the flare as well as
post-flare periods. The WPS obtained from the GOLF data shows post-flare enhancements in the global
velocity oscillations. The corresponding GWPS shows significantly enhanced power beyond 5.5 mHz, the acoustic
cut-off frequency of the solar photosphere. This is different from the high-frequency power
enhancement seen in the GWPS obtained from the full-disk collapsed GONG velocity data
(c.f., right panel of the Figure~7(a)). However, the bands of enhancements seen in the two
cases are different, which is because of the following. The GONG observations will remain
photospheric in nature during the quiet and flaring conditions as Ni~{\sc i}~line is
formed deep in the photosphere. However, GOLF observations are sensitive to photospheric heights
during the quiet phase while it becomes sensitive to chromospheric heights during the
flare \citep{cessa10}. Therefore, the global pseudo-modes observed by the GOLF will have
contributions from chromospheric oscillations during the flare and later to the flare, it
will be sensitive to the photospheric oscillations. The GWPS obtained from the GOLF data
will have mixed effect from both, chromospheric and photospheric oscillations. The Figure~6(b)
shows the scale-average
variance obtained from the GOLF data for the period 01:00~UT to 04:15~UT. This is again a little
different from the scale-average variance obtained from the GONG full-disk collapsed velocity
data (Figure~7(b)) which could be attributed to the same: different heights of formation of
the Na~D and Ni~{\sc i}~lines. In the Figure~6(b),
one can notice a clear difference in the behavior of the scale-average variance before the flare as compared
to its behavior after the flare . We observe strong high-frequency pulses (above 90~$\%$ confidence level)
in the Sun after the flare. This power enhancement is higher than what was seen for a X17 class
flare \citep{kumar10}. However, at this point in time, we can not give any definite physical scenario
to explain this global behavior. In summary, we note that the significant enhancement of power above the
photospheric cut-off frequency ($\sim$5.3~mHz) in the localized velocity oscillations during the flare
appears to be accompanied by the flare related enhancements seen in disk-integrated velocity
observations of the Sun. If confirmed, these results might be indicative of a connection
between flare induced localized seismic response of the solar photosphere and the excitation
of global oscillations in the Sun. However, the future multi-wavelength observations of the
Sun-as-a-star with GOLF-NG \citep{turck08} and SONG \citep{grund11} along with the high-cadence
full-disk data from Solar Dynamics Observatory ({\em SDO}) spacecraft would be very useful for this kind
of study.

The above studies will refine our knowledge about the seismic counterparts of transient events,
such as flares, in the Sun. The recent space missions dedicated to asteroseismology, such as
{\em Kepler} \citep{basr05,chap11} and CoRot \citep{michel08,garc09}, provided high-quality
data to probe the magnetic activity cycles \citep{garc10} and starspots \citep{mosser09,mathur10}
in other stars. Therefore, we can also hope for identifying the astroseismic signature of stellar
flares with better understanding of such connections.

\acknowledgments

This work utilizes data obtained by the Global Oscillation Network Group (GONG) program, managed by
the National Solar Observatory, which is operated by AURA, Inc. under a cooperative agreement with
the National Science Foundation. The data were acquired by instruments operated by the Big Bear
Solar Observatory, High Altitude Observatory, Learmonth Solar Observatory, Udaipur Solar
Observatory, Instituto de Astrof\'{\i}sica de Canarias, and Cerro Tololo Interamerican Observatory.
We also acknowledge the use of data from GOLF instrument on board {\em SOHO} spacecraft. The {\em
SOHO} is a joint mission under cooperative agreement between ESA and NASA. This work has been
partially supported by the CNES/GOLF grant at the Service d'Astrophysique (CEA/Saclay). NCAR is partially
funded by the National Science Foundation. We are very much thankful to the anonymous referee
for the fruitful comments and suggestions which improved the presentation of this work. We are
also thankful to John Leibacher, Frank Hill, H. M. Antia,
Thierry Appourchaux, Paul Cally, Rudolf Komm, Irene Gonz\'alez-Hern\'andez, Richard Bogart and
Sushanta Tripathy for useful discussions related to this work.

\begin{figure*}
\centering
\includegraphics[width=0.411\textwidth]{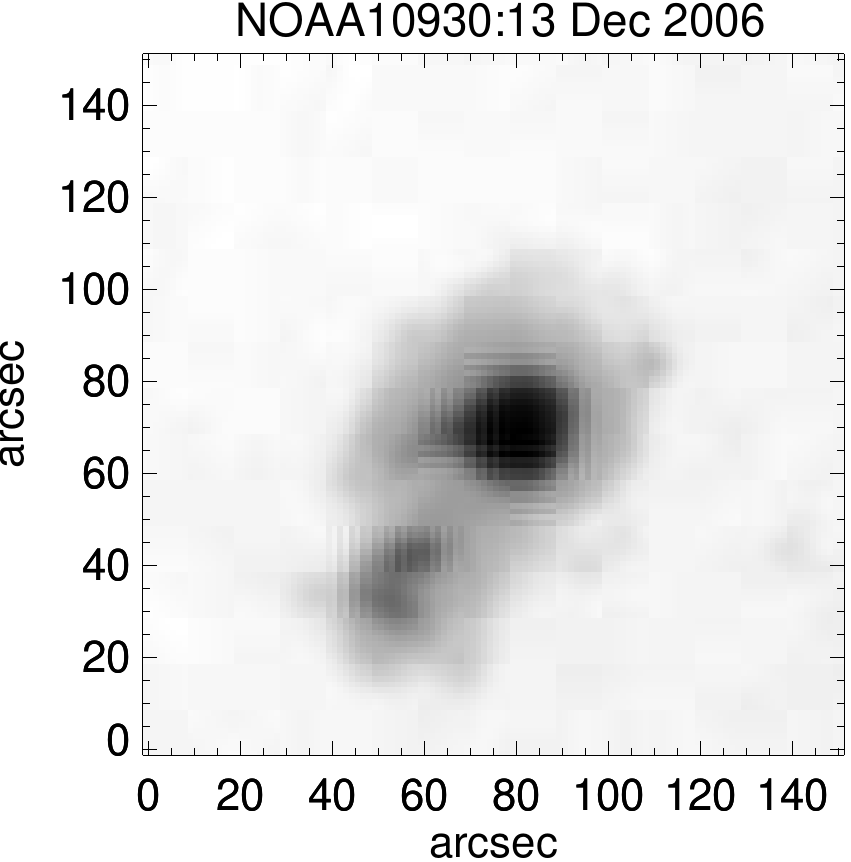} \hspace*{0.34 cm}
\includegraphics[width=0.493\textwidth]{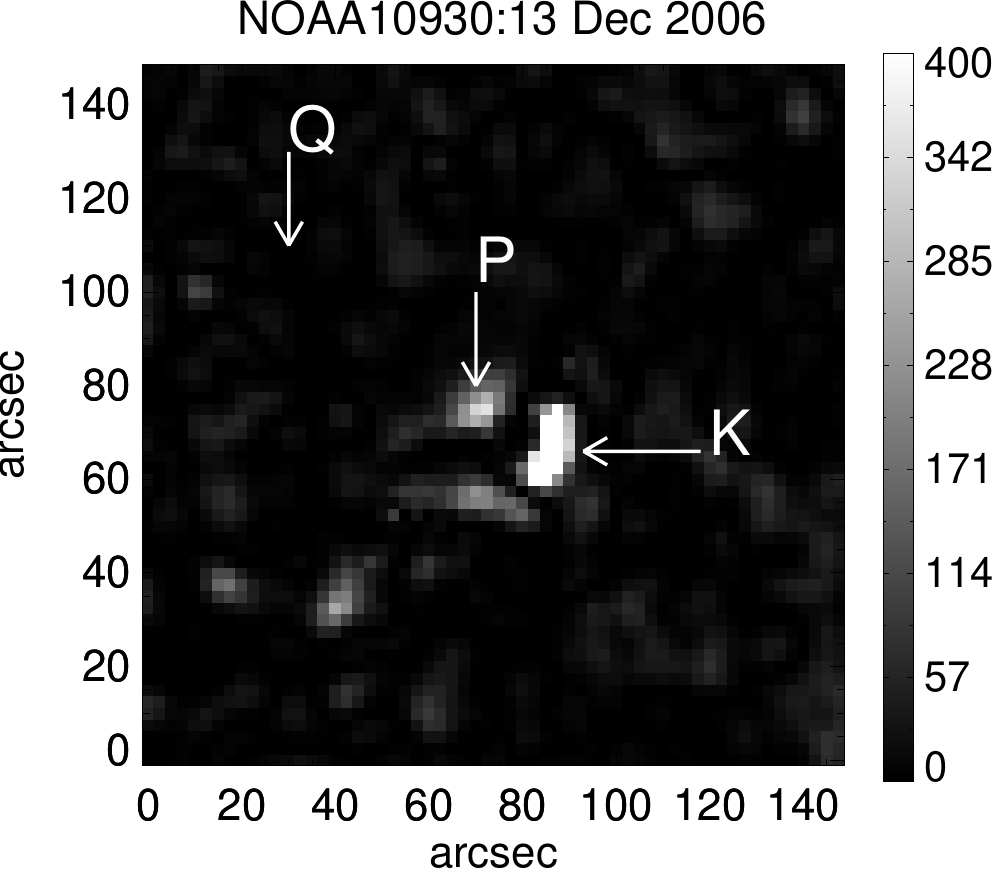}  \\
\vspace*{-.452cm}
\hspace*{0.4cm}(a)
\hspace*{7.0cm}(b)
\includegraphics[width=0.8\textwidth]{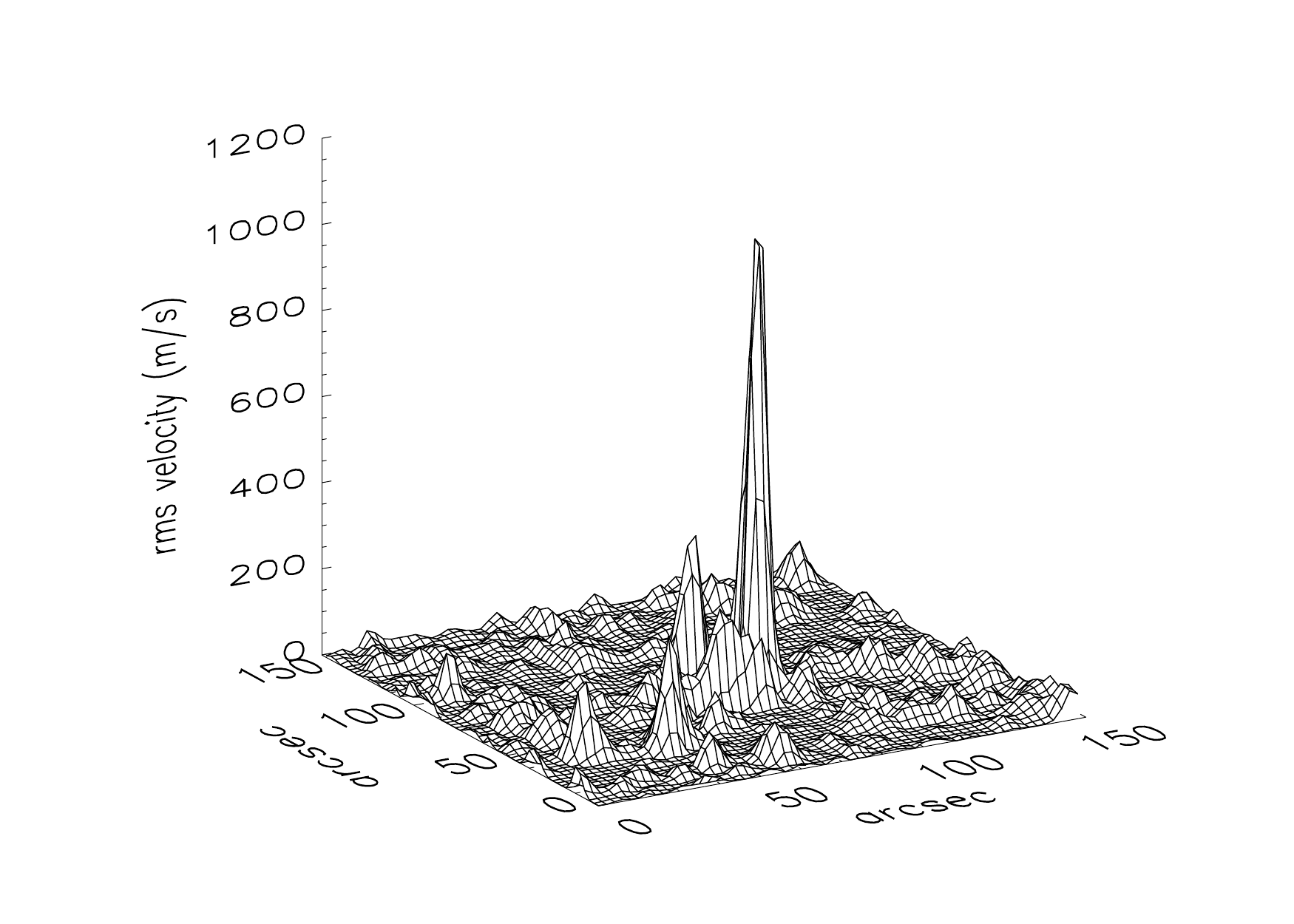}\\
\vspace*{-0.57cm}
\hspace*{2.5cm}(c)
\caption{(a) Continuum intensity map of the active region NOAA 10930 obtained by the GONG
instrument on 13 December 2006 at 02:26~UT. The impulsive phase of the flare was
around 02:25~UT as reported by NoRP. (b) The grid shows the {\em root mean square (rms)}
velocity map computed from the tracked GONG Dopplergrams for the period between 02:20~UT and
03:02~UT (spanning the flare event on 13 December 2006) for the same field-of-view as chosen for
the intensity map shown in (a). The bright patchy region indicated by `K' shows the location
of the suddenly enhanced velocity oscillations in the localized penumbral region of the active
region. `P' shows the location of another penumbral region, diagonally opposite to `K', where the
velocity enhancement is also significant. One of the quiet regions of the Sun is indicated by `Q' in this
image. (c) Surface plot of the {\em (rms)} velocity map as shown in (b). The significant peaks
corresponding to `K' and `P' are clearly visible in this surface plot.}
\end{figure*}

\begin{figure*}
\centering
\includegraphics[width=0.520\textwidth]{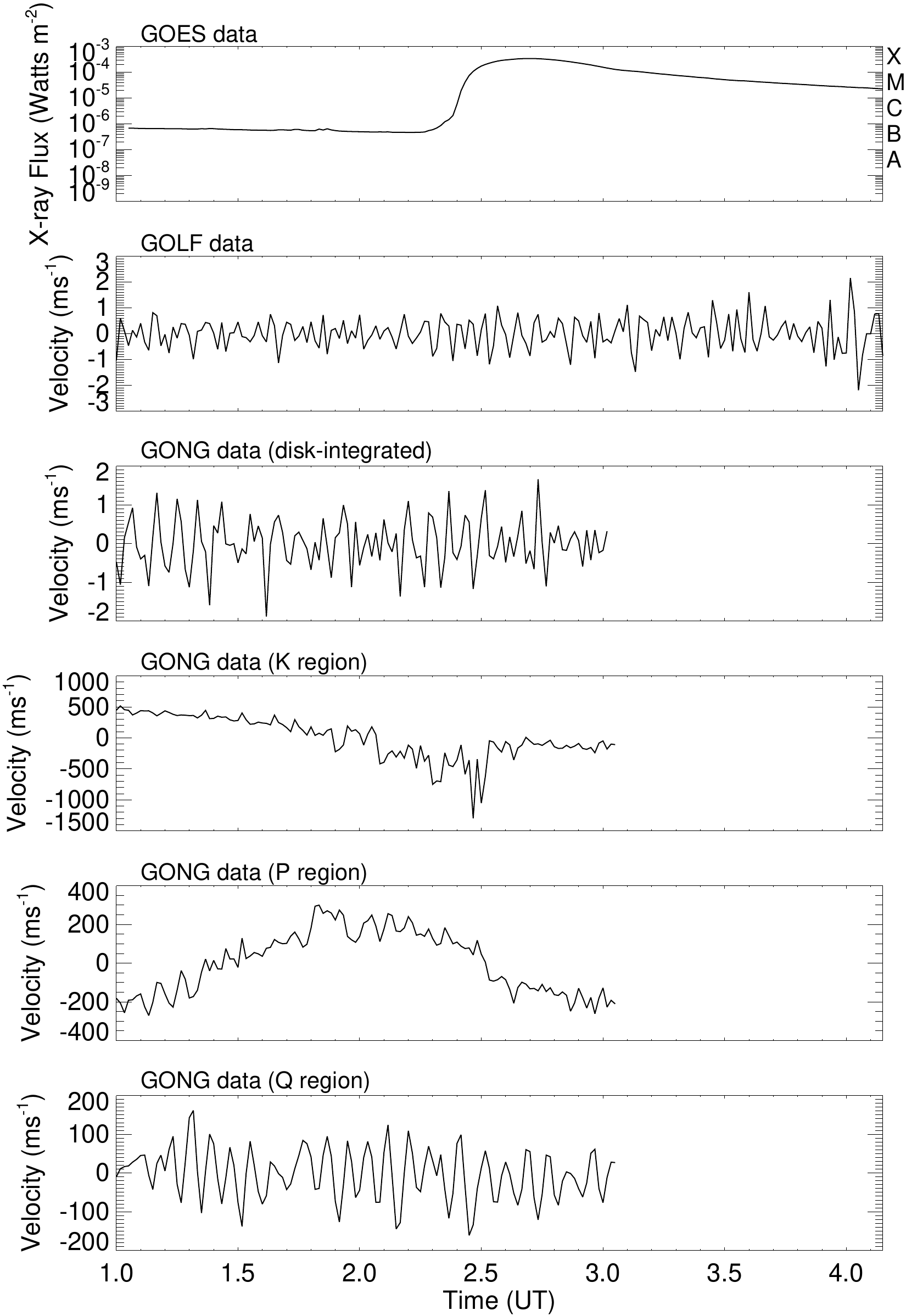}\\
\caption{\label{label} The top panel shows temporal evolution of soft X-ray flux from the Sun as
observed with the {\em GOES} satellite during 01:00-04:15~UT. The flare maximum is seen around
02:35~UT in the {\em GOES} observations. In the second panel from the top, we show the temporal
evolution of disk-integrated velocity observations of the Sun obtained by {\em SOHO}/GOLF instrument
during 01:00-04:15~UT spanning the flare event of 13 December 2006. The third panel from the top
shows the GONG full-disk collapsed velocity signals during 01:00-03:02~UT. The fourth panel from
the top shows the time series of line-of-sight
velocity signals (01:00-03:02~UT) averaged over an area of 3$\times$3 pixels surrounding the centroids
of the bright patchy penumbral region (K) as detected in the {\em rms} velocity map computed from the GONG
Dopplergrams and shown in the Figure~1(b). In the fifth and sixth panels from the top, we show the
similar velocity time series for the penumbral region `K' and the quiet region `Q', respectively,
as shown in the Figure~1(b). Here, the fourth and sixth panels have been adopted from the Figure~2
and Figure~3 in \cite{kumar11} by permission.}
\end{figure*}

\begin{figure*}
%\begin{tabular}{cc}
\centering
\includegraphics[width=0.407\textwidth]{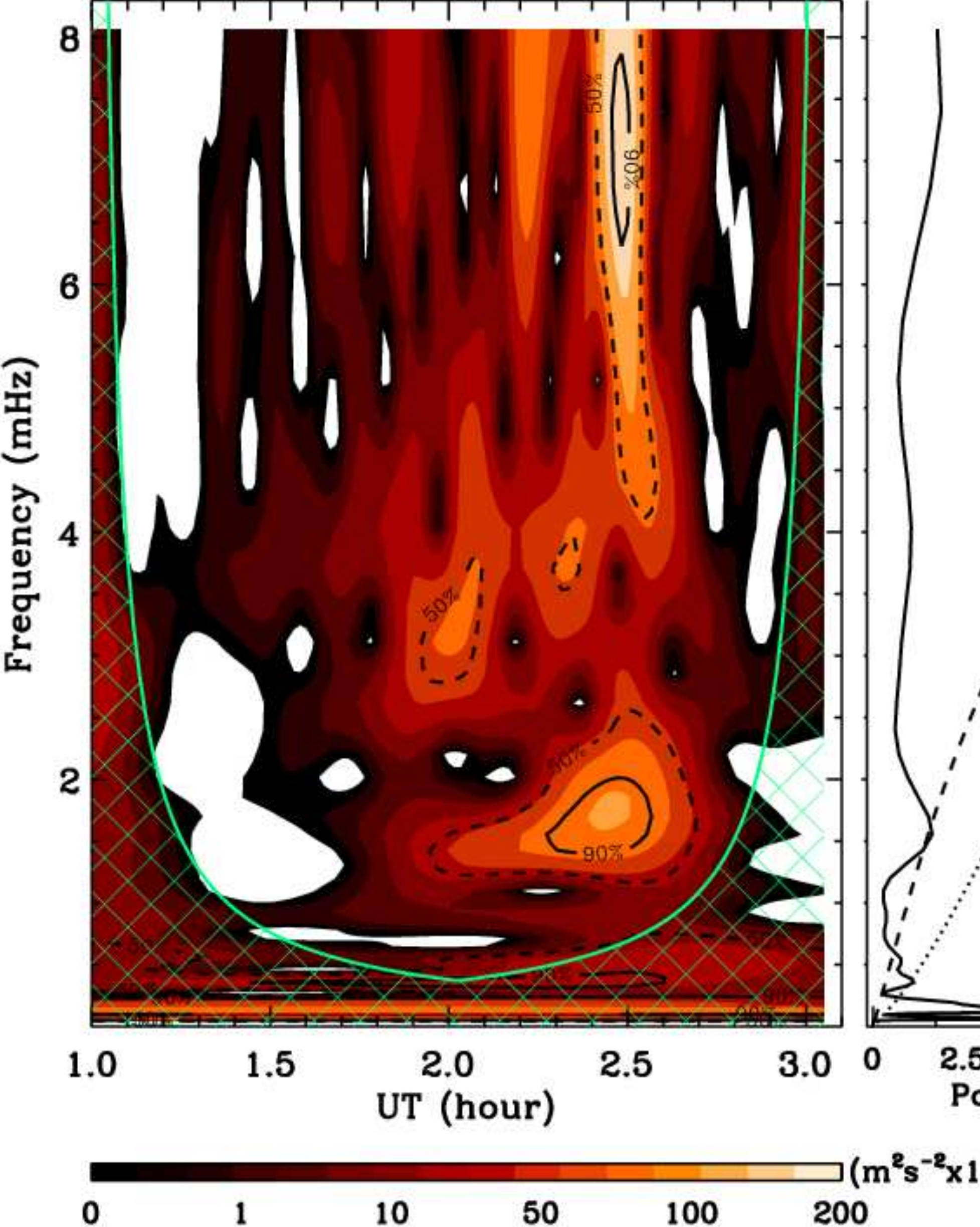} \hspace*{2.5 cm}
\includegraphics[width=0.407\textwidth]{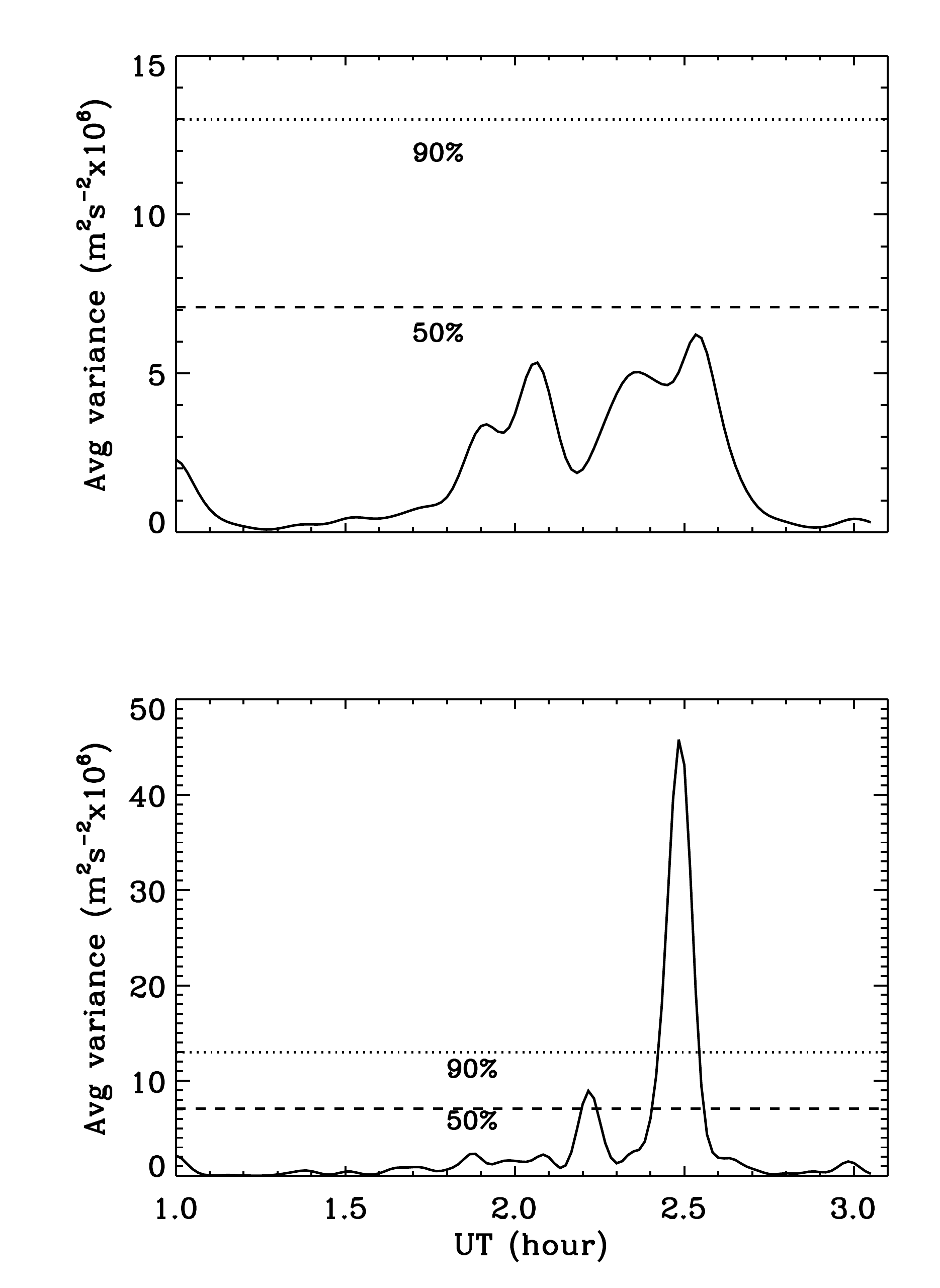}\\
\hspace*{0.4cm}(a)
\hspace*{7.0cm}(b)
\caption{\label{label}(a) The left panel shows the Wavelet Power Spectrum (WPS) computed from the
time series of line-of-sight velocity signals as shown for the penumbral region `K' in the Figure~2
(fourth panel from the top). The right panel shows the Global-wavelet Power Spectrum (GWPS) computed from
this time series. In the WPS, the solid lines correspond to regions with 90\%
confidence level whereas the dashed lines are for 50\% confidence level and the hatched region indicates
the cone of influence. In the GWPS, the dashed line is for 50\% significance level and the dotted line
is for 90\% significance level. (b) The plots illustrate scale-average time series for the WPS in the
frequency regime: 2--5~mHz (top panel), and 5--8~mHz (bottom panel). In these plots, the
dotted line corresponds to 90\% significance level and the dashed line corresponds to 50\% significance
level. These plots are adapted version of the Figure~2 in \cite{kumar11} by permission.}
%\end{tabular}
\end{figure*}

\begin{figure*}
\centering
\includegraphics[width=0.407\textwidth]{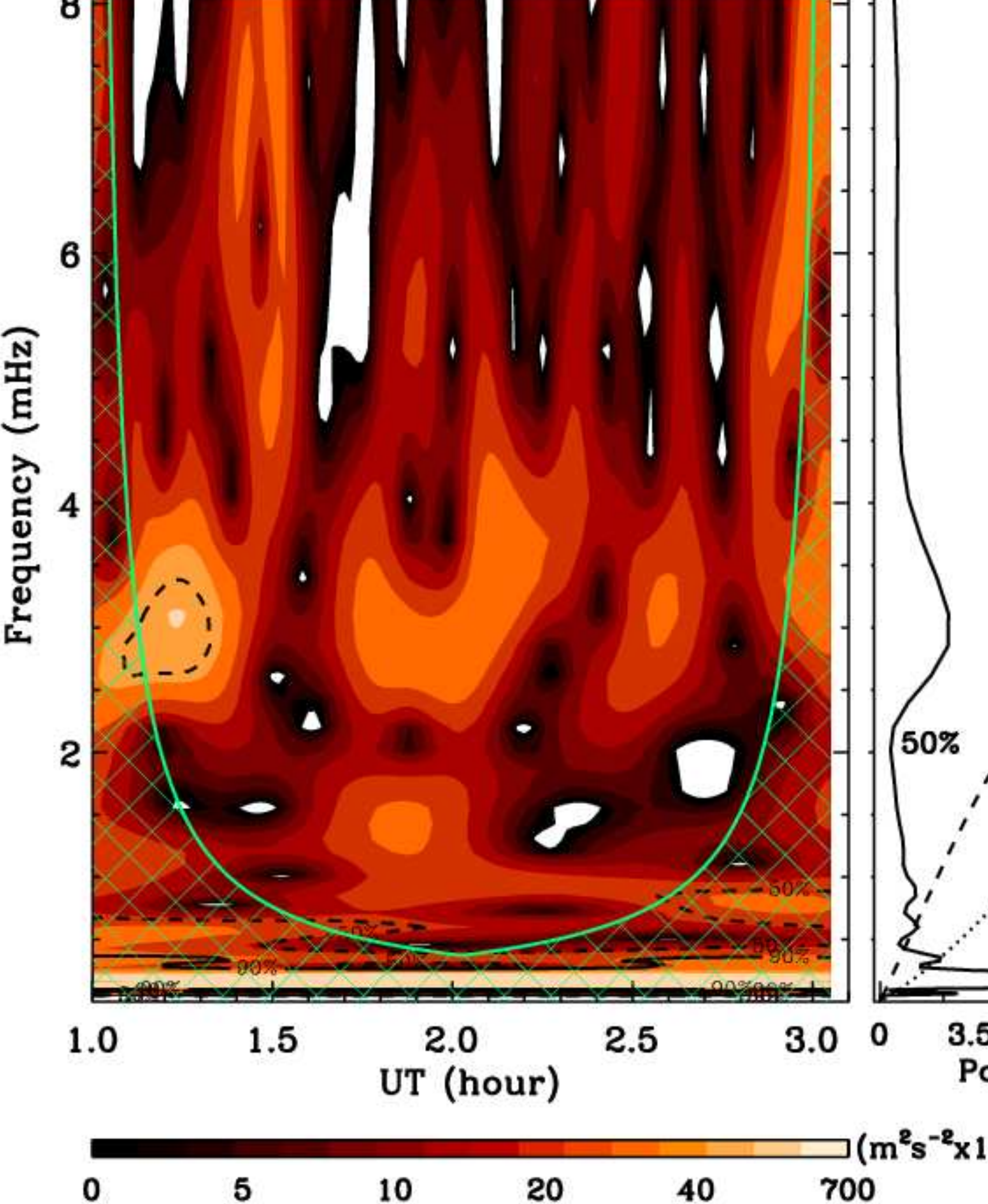} \hspace*{2.5 cm}
\includegraphics[width=0.407\textwidth]{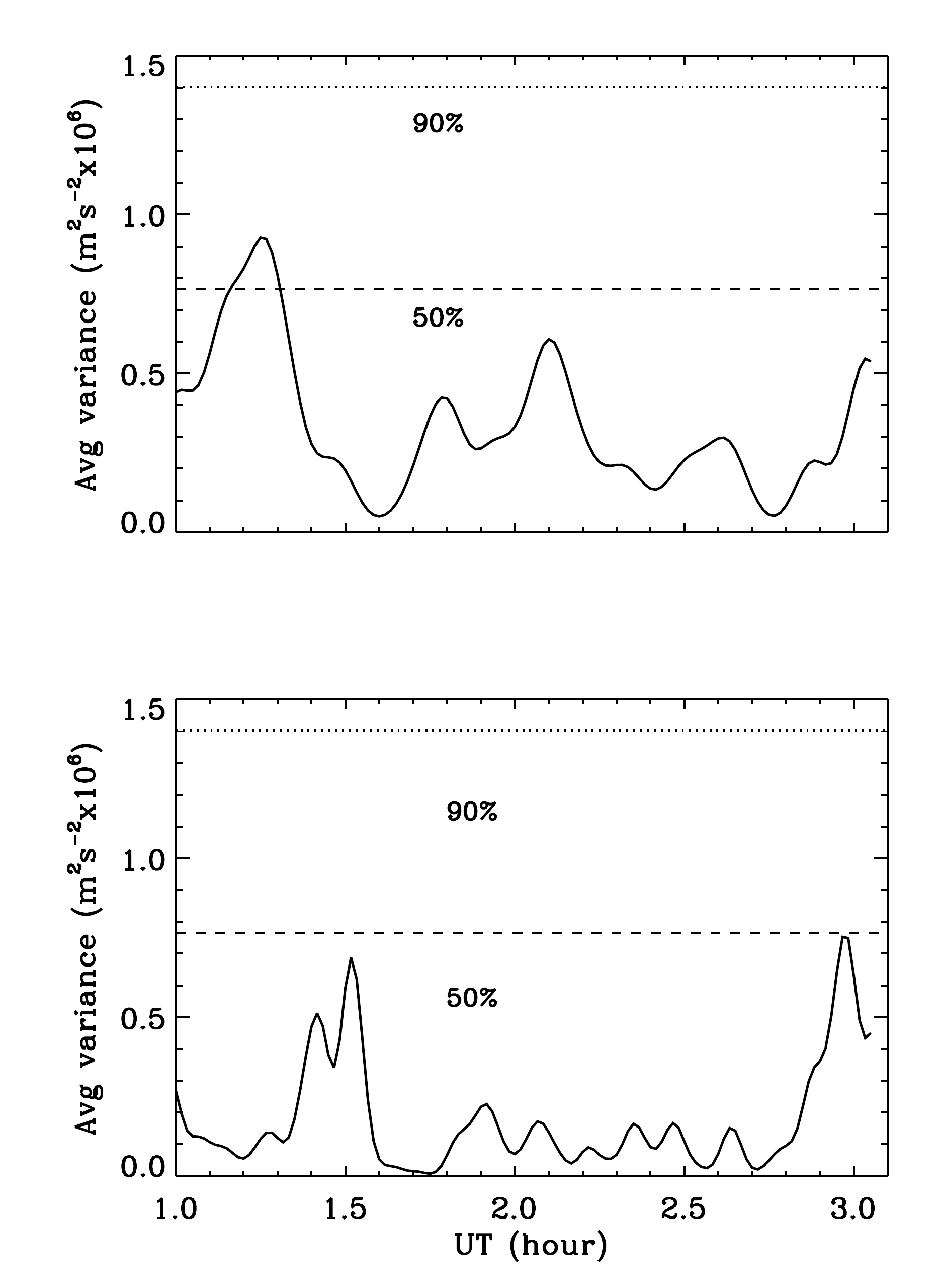}\\
\hspace*{0.4cm}(a)
\hspace*{7.0cm}(b)
\caption{\label{label} Same as Figure~3, but for the velocity time series from the penumbral region `P'
as shown in the Figure~2 (fifth panel from the top). The region `P' does not show any transient velocity signals
during the flare. The normal $p$ modes also appear to be suppressed due to high magnetic field concentration.}
\end{figure*}

\begin{figure*}
\centering
\includegraphics[width=0.407\textwidth]{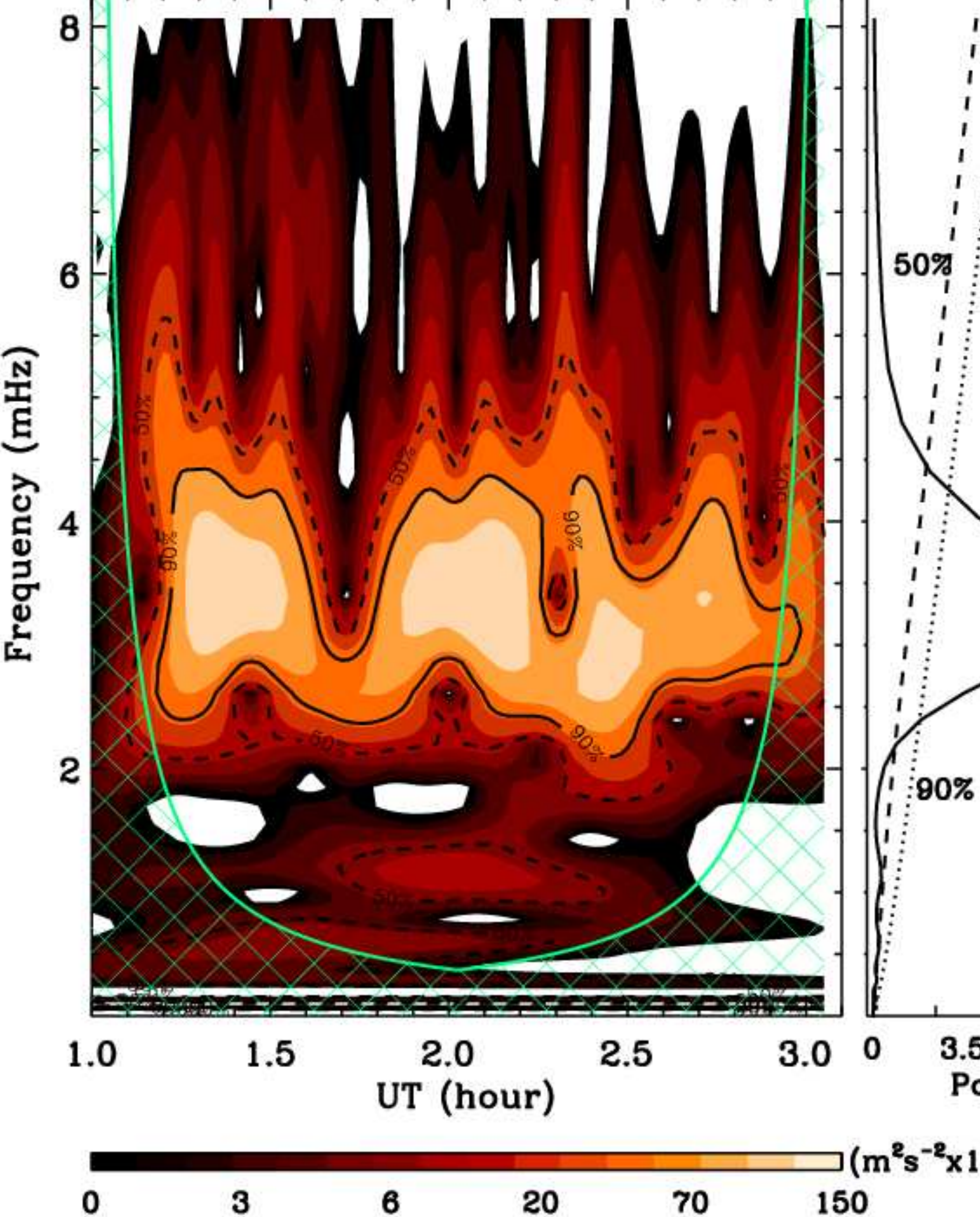} \hspace*{2.5 cm}
\includegraphics[width=0.407\textwidth]{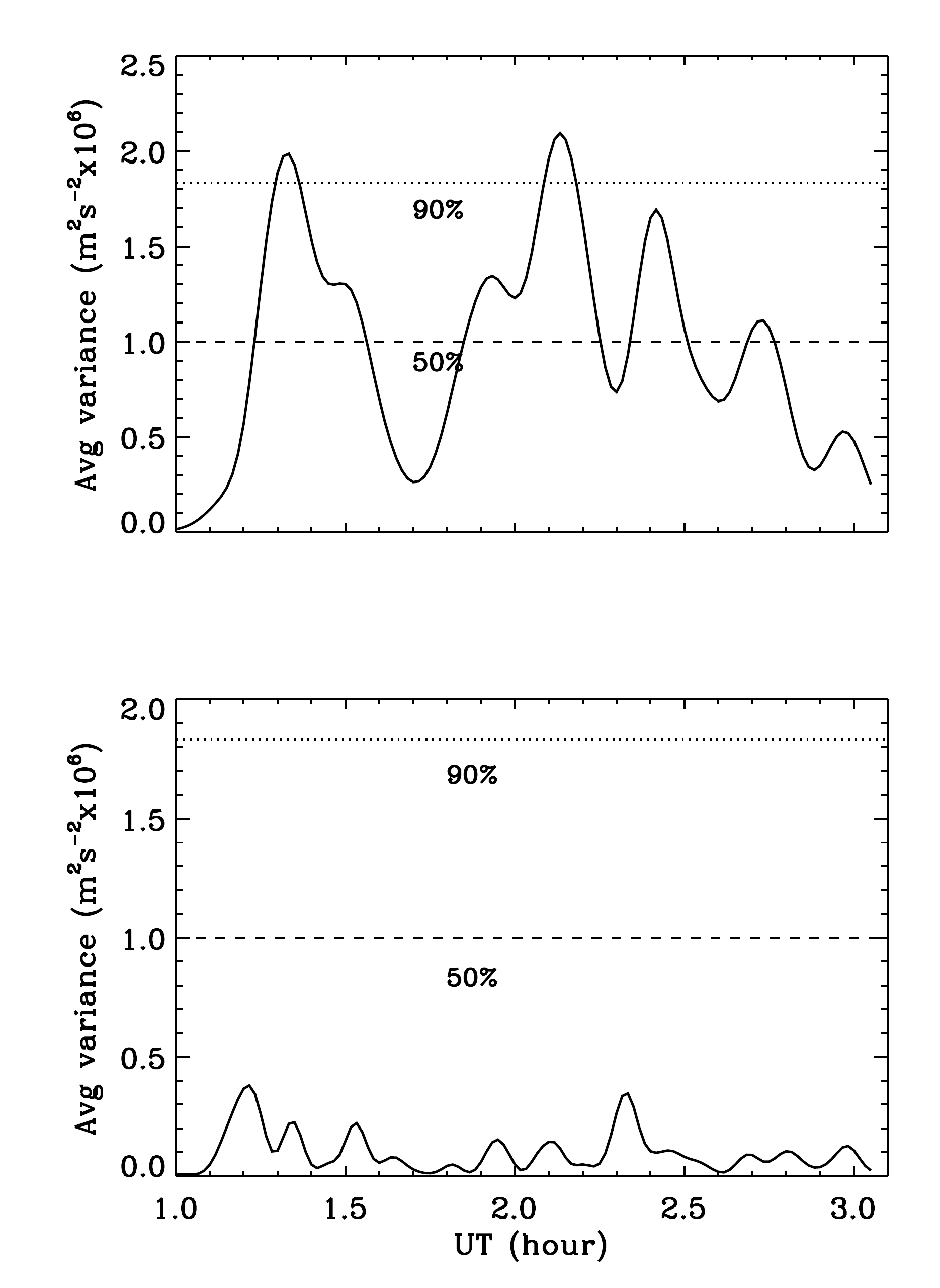}\\
\hspace*{0.4cm}(a)
\hspace*{7.0cm}(b)
\caption{\label{label} Same as Figure~3, but for the velocity time series from the quiet region `Q' as
shown in the Figure~2 (sixth panel from the top). Here, the dominant 5-min oscillations are clearly seen
in the Wavelet Power Spectrum, Global-wavelet Power Spectrum and the scale-average time series. These
plots are adapted version of the Figure~3 in \cite{kumar11} by permission.}
\end{figure*}

\begin{figure*}
\centering
\includegraphics[width=0.407\textwidth]{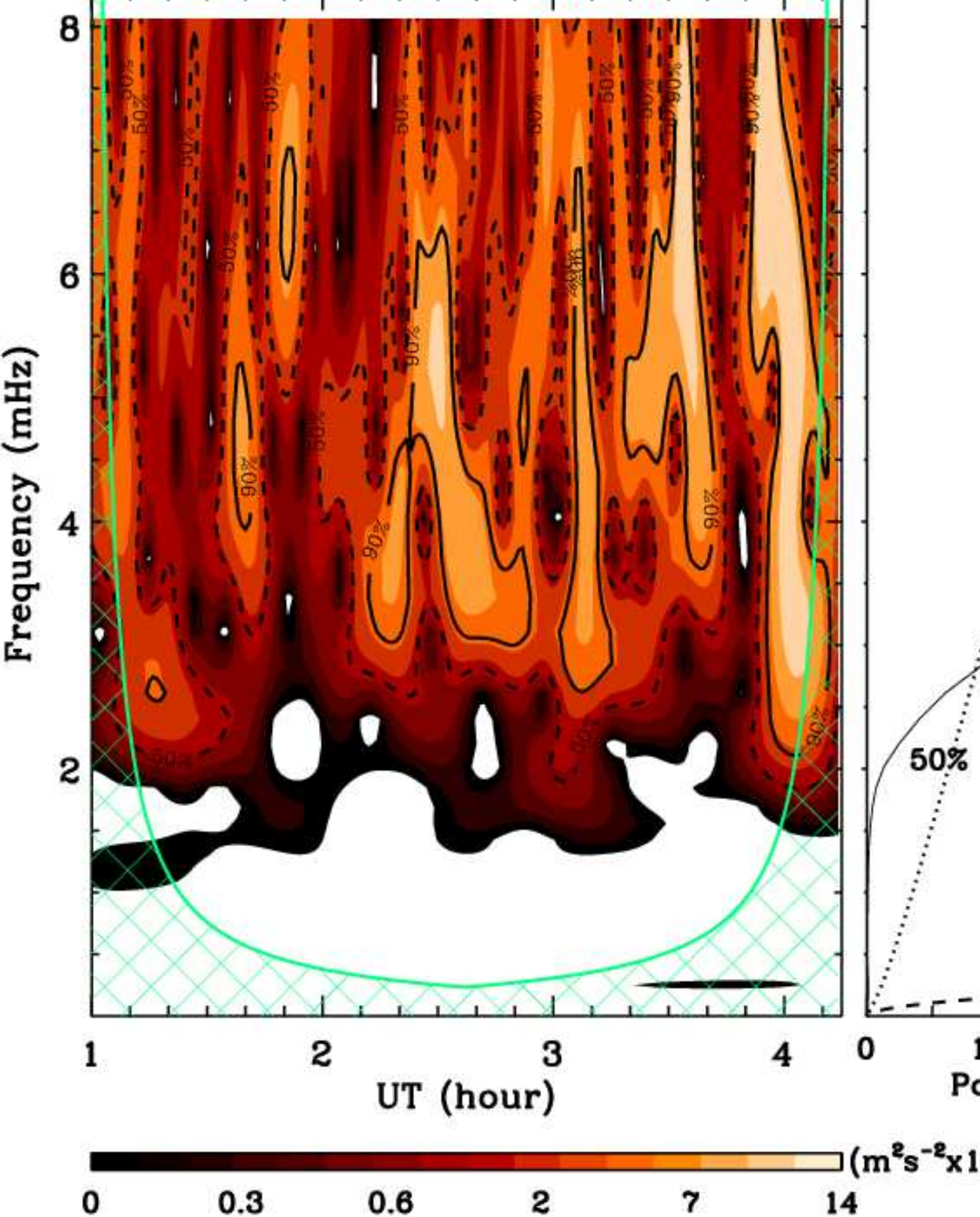} \hspace*{2.5 cm}
\includegraphics[width=0.407\textwidth]{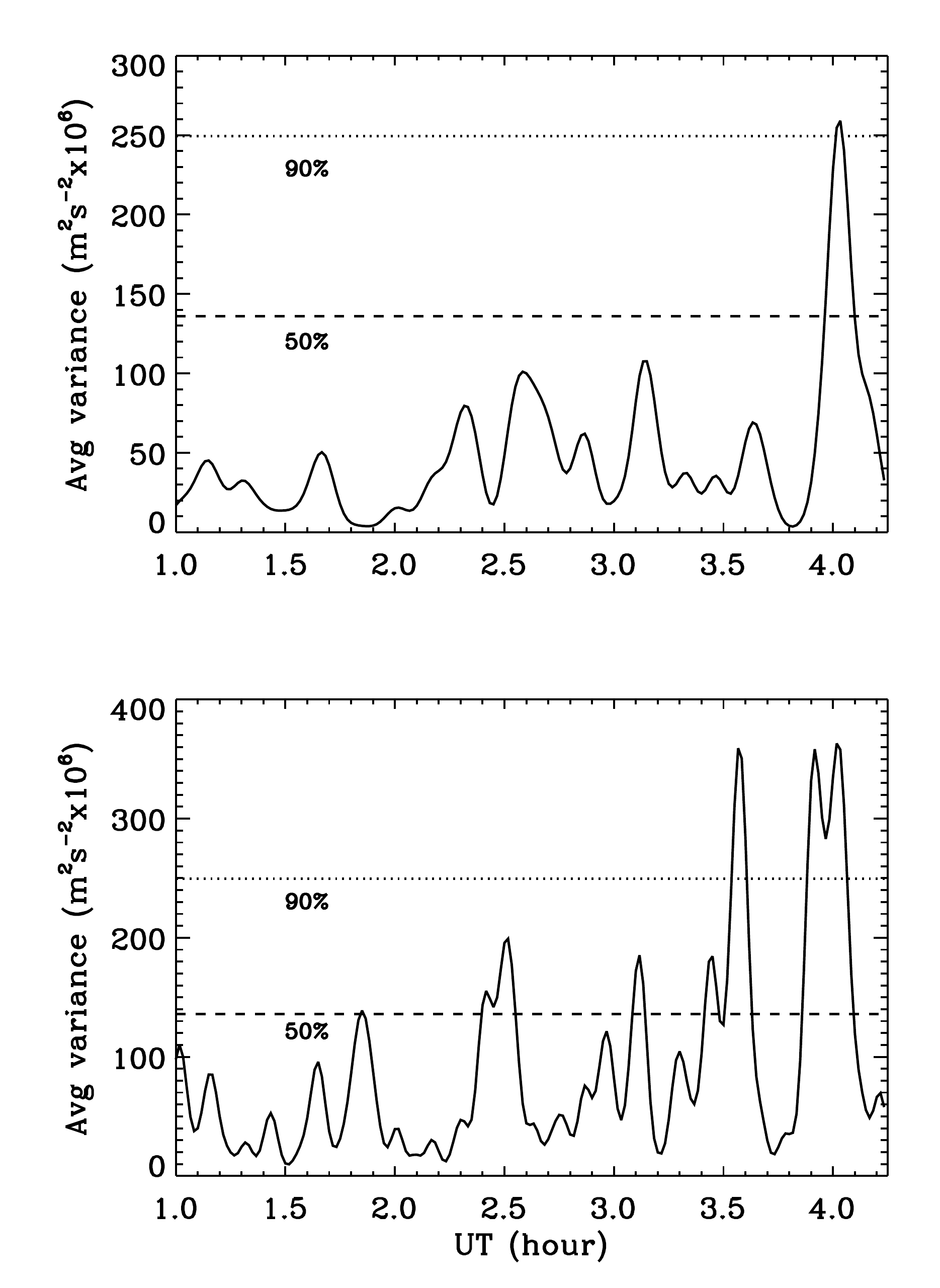}\\
\hspace*{0.4cm}(a)
\hspace*{7.0cm}(b)
\caption{\label{label}(a) The left panels show the Wavelet Power Spectrum computed from the time
series of disk-integrated velocity observations (as shown in the second panel from the top in the Figure~2)
of the Sun obtained by {\em SOHO}/GOLF instrument during 01:00-04:15~UT spanning the flare event of
13 December 2006. The right panel shows the Global-wavelet Power Spectrum computed from this time series.
(b) Same as Figure~3, but using GOLF data.}
\end{figure*}

\begin{figure*}
\centering
\includegraphics[width=0.49\textwidth]{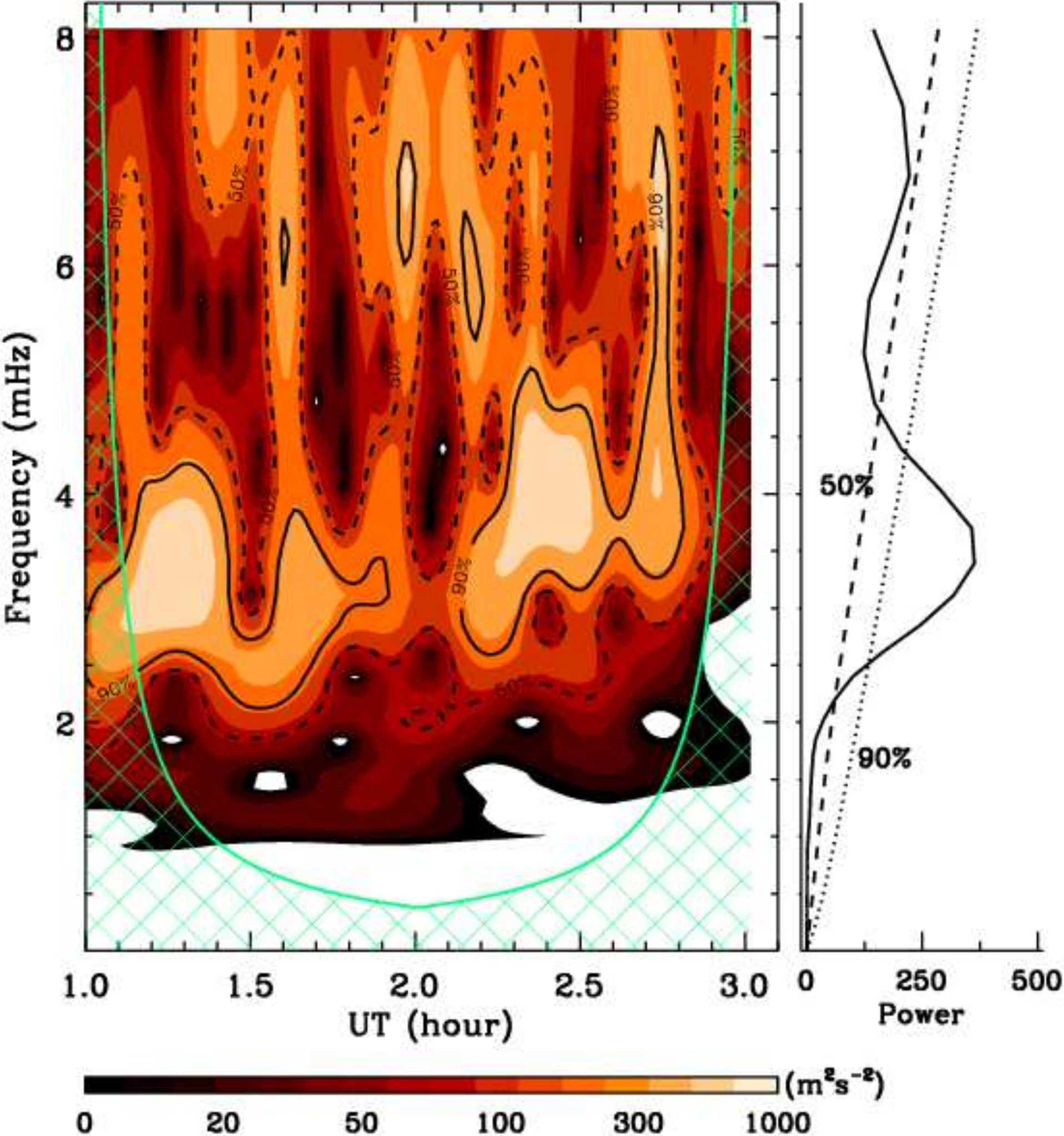} \hspace*{.2 cm}
\includegraphics[width=0.41\textwidth]{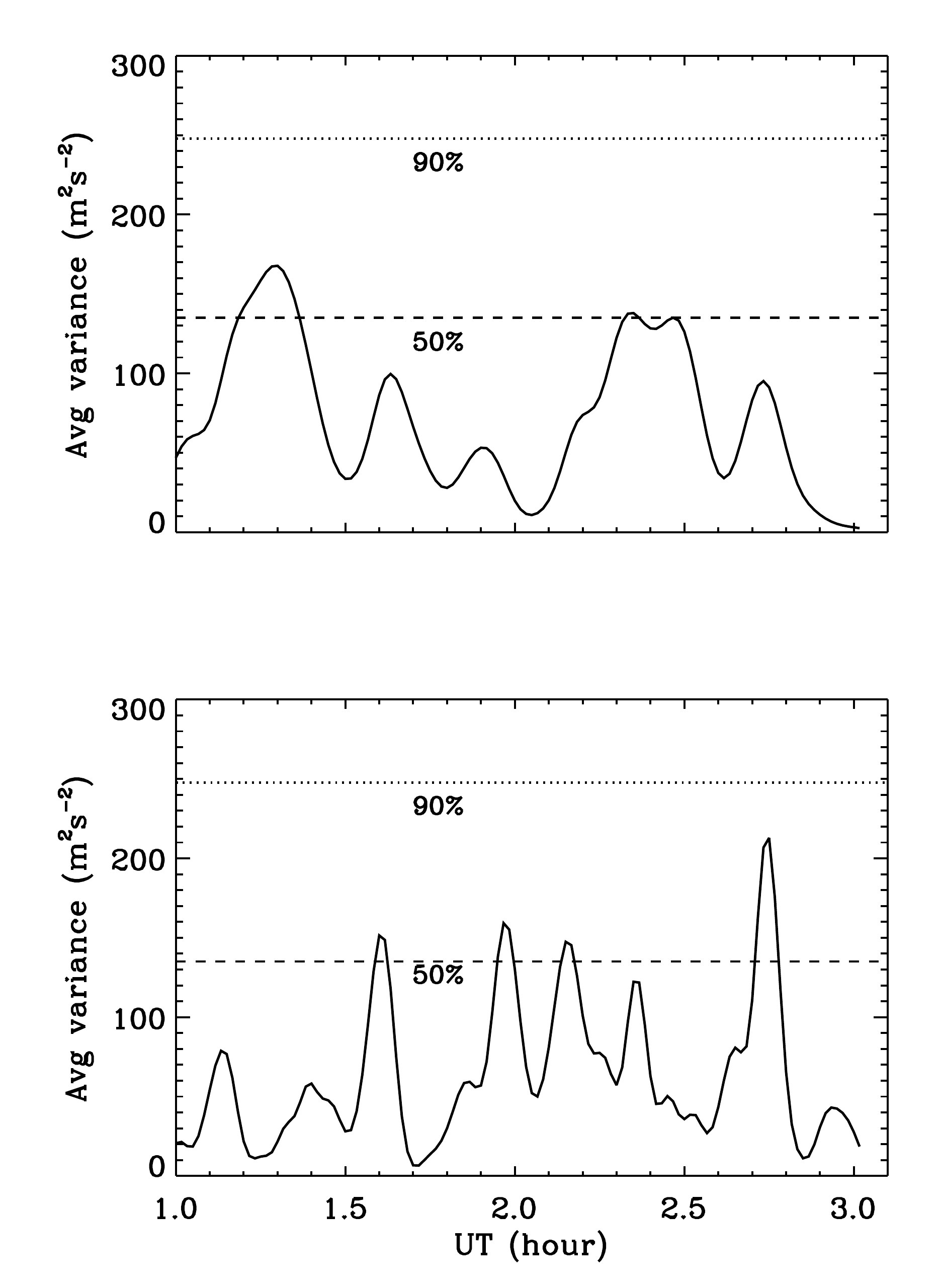}\\
\hspace*{0.4cm}(a)
\hspace*{7.0cm}(b)
\caption{\label{label}(a) The left panels show the Wavelet Power Spectrum computed from the time
series of GONG full-disk collapsed velocity observations (as shown in the third panel from the
top in the Figure~2) during 01:00-03:02~UT spanning the flare event of 13 December 2006. The right
panel shows the Global-wavelet Power Spectrum computed from this time series.
(b) Same as Figure~3, but using GONG full-disk collapsed velocity data.}
\end{figure*}

\end{document}